\documentstyle[preprint,array,aps]{revtex}
\begin{document}
\draft
\title
{Relativistic entanglement of quantum states and nonlocality of
Einstein-Podolsky-Rosen(EPR) paradox}
\author{Doyeol Ahn$^{1,2}${\footnote{e-mail:dahn@uoscc.uos.ac.kr}},
Hyuk-jae Lee$^1${\footnote{e-mail:lhjae@iquips.uos.ac.kr}}and Sung
Woo Hwang$^{1,3}${\footnote{e-mail:swhwang@korea.ac.kr}}}
\address{
$^1$Institute of Quantum Information Processing and Systems,
University of Seoul, Seoul, 130-743, Korea\\
$^2$Department of Electrical and computer Engineering, University
of Seoul,
Seoul, 130-743, Korea\\
$^3$Department of Electronic Engineering, Korea University, Seoul,
136-701, Korea}

\maketitle


\baselineskip24pt



\begin{abstract}
Relativistic bipartite entangled quantum states is studied to show
that Nature doesn't favor nonlocality for massive particles in the
ultra-relativistic limit. We found that to an observer (Bob) in a
moving frame $S'$, the entangled Bell state shared by Alice and
Bob appears as the superposition of the Bell bases in the frame
$S'$ due to the requirement of the special relativity. It is shown that
the entangled pair satisfies the Bell's inequality when the boost
speed approaches the speed of light, thus providing a counter
example for nonlocality of Einstein-Podolsky-Rosen(EPR) paradox.
\end{abstract}
\vspace{.25in}



Entanglement of bipartite quantum states is of fundamental
interest for quantum information processing such as quantum
computation\cite{benn}--\cite{eker},
teleportation\cite{bras}--\cite{lee} and clock
synchronization\cite{chua}--\cite{hwan}. How does the entangled
quantum states appear to an observer in a different Lorentz frame
would be an interesting question, potentially related to the clock
synchronization problem. Another unsolved problem, perhaps more
important than the above one, is the violation of the local
causality in quantum mechanics by measurement process, so called,
the Einstein-Podolsky-Rosen(EPR) paradox\cite{eins} and the Bell's
theorem\cite{bell}, suggesting the existence of an instantaneous
action between distant measurements. This subtle question still
remains to be answered even though there have been several
works\cite{ahar}--\cite{peres} relating the relativity, entropy,
entanglements and the quantum operations. Recently, Czachor\cite{czac}
suggested that the degree of violation of the Bell inequality
depends on the velocity of the pair of spin-$\frac{1}{2}$
particles with respect to the laboratory and it would be interesting to
study the Bell's inequality of relativistically entangled quantum states.

One of the conceptual barriers for the relativistic treatment of
quantum information processing is the difference of the role
played by the wave fields and the state vectors in the quantum
field theory. In non-relativistic quantum mechanics both the wave
function and the state vector in Hilbert space give the
probability amplitude which can be used to define conserved
positive probability densities or density matrix. On the other
hand, in relativistic quantum field theory, the wave fields are
not probability amplitude at all, but operators which create or
destroy particles in spanned by states defined as containing
definite numbers of particles or antiparticles in each normal
mode\cite{wein}. Moreover, there has been a
quandary\cite{ahar}--\cite{pere} whether the quantum states are
Lorentz covariant but according to Weinberg\cite{wein}, the
quantum states viewed from different reference frames can be
represented by the Lorentz transformation.

More recently, Alsing and Milburn\cite{alsi} studied the Lorentz
invariance of entanglement and showed that the entanglement
fidelity of the bipartite state is preserved explicitly. To the
best of our knowledge, their work is the first detailed
calculation of the relativistic quantum entanglement of bipartite
state. However, in their approach, it is not quite clear whether
the entanglement is for the quantum state or the quantum fields
because they started from the entanglement between the $4$-spinors
for the Dirac field. In quantum field theory, the role of the
field is to make the interaction or the S-matrix satisfying the
Lorentz invariance and the cluster decomposition principle. On the
other hand, the information of the particle states is contained in
the state vectors of the Hilbert space spanned by states
containing $0, 1, 2, \cdots,$ particles as in the case of
non-relativistic quantum mechanics\cite{wein}.

In this article, we study the Lorentz transformation properties of
entanglement of bipartite quantum states in the Hilbert space and
provide the counter example for the nonlocality of the EPR
paradox. Throughout the article, we follow Weinberg's
notation\cite{wein}. A multiparticle state vector is denoted by
\begin{equation}
\Psi_{p_1,\sigma_1;p_2,\sigma_2;\cdots}=
a^+(\vec{p_1},\sigma_1)a^+(\vec{p_2},\sigma_2)\cdots\Psi_0,
\label{st}
\end{equation}
where $p_i$ labels the four-momentum, $\sigma_i$ is the spin $z$
component, $a^+(\vec{p_i},\sigma_i)$ is the creation operator
which adds a particle with momentum $\vec{p_i}$, and spin
$\sigma_i$, and $\Psi_0$ is the Lorentz invariant vacuum state.
The Lorentz transformation $\Lambda$ induces a unitary
transformation on vectors in the physical Hilbert space
\begin{equation}
\Psi\rightarrow U(\Lambda)\Psi, \label{ust}
\end{equation}
and the operators $U$ satisfies the composition rule
\begin{equation}
U(\bar{\Lambda})U(\Lambda)=U(\bar{\Lambda}\Lambda), \label{utran}
\end{equation}
while the creation operator has the following transformation
rule\cite{wein}
\begin{equation}
U(\Lambda)a^+(\vec{p},\sigma)U^{-1}(\Lambda)=\sqrt{\frac{(\Lambda
p)^0}{p^0}}\sum_{\bar{\sigma}}{\cal
D}^{(j)}_{\bar{\sigma}\sigma}(W(\Lambda,p))a^+(\vec{p_{\Lambda}},\bar{\sigma}).
\label{atran}
\end{equation}
Here, $W(\Lambda,p)$ is the Wigner's little group element given by
\begin{equation}
W(\Lambda,p)=L^{-1}(\Lambda p)\Lambda L(p), \label{wlamd}
\end{equation}
${\cal D}^{(j)}(W)$ the representation of $W$ for spin $j$,
$p^{\mu}=(\vec{p}, p^0)$ and $(\Lambda
p)^{\mu}=(\vec{p_{\Lambda}}, (\Lambda p)^0)$ with $\mu=1,2,3,0$,
and $L(p)$ is the standard Lorentz transformation such that
\begin{equation}
p^{\mu}=L^{\mu}_{{}\nu}(p)k^{\nu}, \label{sloren}
\end{equation}
where $k^{\nu} =(0,0,0,m)$ is the four-momentum taken in the
particle's rest frame.

The relativistic momentum-conserved entangled Bell states for spin
$\frac{1}{2}$ particles in the rest frame $S$ are defined by
\begin{mathletters}
\begin{eqnarray}
\Psi_{00}&=&\frac{1}{\sqrt{2}}\{
a^+(\vec{p},{{1}\over {2}})a^+(-\vec{p},\frac{1}{2})+a^+(\vec{p},-\frac{1}{2})a^+(-\vec{p},-\frac{1}{2})\}\Psi_0,\label{bello}\\
\Psi_{01}&=&\frac{1}{\sqrt{2}}\{
a^+(\vec{p},\frac{1}{2})a^+(-\vec{p},\frac{1}{2})-a^+(\vec{p},-\frac{1}{2})a^+(-\vec{p},-\frac{1}{2})\}\Psi_0,\label{bellt}\\
\Psi_{10}&=&\frac{1}{\sqrt{2}}\{
a^+(\vec{p},\frac{1}{2})a^+(-\vec{p},-\frac{1}{2})+a^+(\vec{p},-\frac{1}{2})a^+(-\vec{p},\frac{1}{2})\}\Psi_0,\label{bellth}\\
\Psi^{11}&=&\frac{1}{\sqrt{2}}\{
a^+(\vec{p},\frac{1}{2})a^+(-\vec{p},-\frac{1}{2})-a^+(\vec{p},-\frac{1}{2})a^+(-\vec{p},\frac{1}{2})\}\Psi_0,\label{bellf}\\
\nonumber
\end{eqnarray}
\end{mathletters}
where $\Psi_0$ is the Lorentz invariant vacuum state. It is
straightforward to see that the momentum-conserved Bell states
$(\ref{bello})-(\ref{bellf})$ have both the space inversion ($\cal
P$) and the time-reversal ($\cal T$) symmetries.

For an observer in another reference frame $S'$ described by an
arbitrary boost $\Lambda$, the transformed Bell states are given
by
\begin{equation}
\Psi_{ij}\rightarrow U(\Lambda)\Psi_{ij}. \label{ptrans}
\end{equation}

For example, from equations (\ref{atran}) and (\ref{bello}),
$U(\Lambda)\Psi_{00}$ becomes
\begin{eqnarray}
U(\Lambda)\Psi_{00}&=&\frac{1}{\sqrt{2}}\{
U(\Lambda)a^+(\vec{p},\frac{1}{2})U^{-1}(\Lambda)U(\Lambda)a^+(-\vec{p},\frac{1}{2})U^{-1}(\Lambda)\nonumber\\
& &+
U(\Lambda)a^+(\vec{p},-\frac{1}{2})U^{-1}(\Lambda)U(\Lambda)a^+(-\vec{p},-\frac{1}{2})U^{-1}(\Lambda)\}U(\Lambda)\Psi_0
\nonumber\\
&=&\frac{1}{\sqrt{2}}\sum_{\sigma,\sigma'}\{\sqrt{\frac{(\Lambda
p)^0}{p^0}}{\cal D}^{(\frac{1}{2})}_{\sigma\frac{1}{2}}(W(\Lambda,
p))\sqrt{\frac{(\Lambda {\cal P}p)^0}{({\cal P}p)^0}}{\cal
D}^{(\frac{1}{2})}_{\sigma'\frac{1}{2}}(W(\Lambda, {\cal
P}p))a^+(\vec{p}_{\Lambda},\sigma)a^+(\vec{p}_{\Lambda},\sigma')\nonumber\\
&&+\sqrt{\frac{(\Lambda p)^0}{p^0}}{\cal
D}^{(\frac{1}{2})}_{\sigma-\frac{1}{2}}(W(\Lambda,
p))\sqrt{\frac{(\Lambda {\cal P}p)^0}{({\cal P}p)^0}}{\cal
D}^{(\frac{1}{2})}_{\sigma'-\frac{1}{2}}(W(\Lambda, {\cal
P}p))a^+(\vec{p}_{\Lambda},\sigma)a^+(\vec{p}_{\Lambda},\sigma')\}\Psi_0\label{tbell}
\end{eqnarray}
and so on. Here ${\cal P}$ is the space-inversion operator. For
simplicity, we assume that $\vec{p}$ is in $z$-direction,
$\vec{p}=(0,0,p)$ and the boost $\Lambda$ is in $x$-direction.
Then, we have
\begin{mathletters}
\begin{eqnarray}
L(p)&=&\left[\begin{array}{cccc}
                 1  &  0   &  0        &  0 \\
                 0  &  1   &  0        &  0 \\
                 0  &  0   & \cosh\eta &\sinh\eta \\
                 0  &  0   & \sinh\eta &\cosh\eta
             \end{array} \right],
\label{lp}\\
L({\cal P}p)&=&\left[\begin{array}{cccc}
                 1    &  0    &  0        &  0\\
                 0    &  1    &  0        &  0\\
                 0    &  0    & \cosh\eta &-\sinh\eta\\
                 0    &  0    &-\sinh\eta &\cosh\eta
             \end{array} \right],
\label{lpp}
\end{eqnarray}
and
\begin{equation}
\Lambda=\left[\begin{array}{cccc}
                 \cosh\omega  &  0    &  0        &  \sinh\omega\\
                 0            &  1    &  0        &  0          \\
                 0            &  0    &  1        &  0          \\
                 \sinh\eta    &  0    &  0        &\cosh\omega
             \end{array} \right]
\label{llp}
\end{equation}
\end{mathletters}
where $\eta$ and $\omega$ are the boost in $z$- and
$x$-directions, respectively. The matrix representation of the
Wigner's little group $W$ is given by\cite{halp}
\begin{mathletters}
\begin{equation}
{\cal D}^{(\frac{1}{2})}_{\sigma'\sigma}(W(\Lambda,p))
=\left(\begin{array}{cc}
                      \cos(\frac{\Omega_p}{2}) & -\sin(\frac{\Omega_p}{2})\\
                      \sin(\frac{\Omega_p}{2}) &  \cos(\frac{\Omega_p}{2})
                      \end{array}\right),
\label{rot}
\end{equation}
and
\begin{equation}
{\cal D}^{(\frac{1}{2})}_{\sigma'\sigma}(W(\Lambda,{\cal P}p))
=\left(\begin{array}{cc}
                      \cos(\frac{\Omega_p}{2}) &  \sin(\frac{\Omega_p}{2})\\
                      -\sin(\frac{\Omega_p}{2}) &  \cos(\frac{\Omega_p}{2})
                      \end{array}\right),
\label{prot}
\end{equation}
\end{mathletters}
where the Wigner angle $\Omega_p$ is defined by
\begin{equation}
\tan\Omega_p = \frac{\sinh\eta \sinh\omega}{\cosh\eta +
\cosh\omega}. \label{htan}
\end{equation}
By substituting equations (\ref{rot}) and (\ref{prot}) into
equation (\ref{tbell}), we obtain
\begin{mathletters}
\begin{eqnarray}
U(\Lambda)\Psi_{00}&=&\frac{(\Lambda p)^0}{p^0}\cos\Omega_p
\frac{1}{\sqrt{2}}\{ a^+(\vec{p},{{1}\over
{2}})a^+(-\vec{p},\frac{1}{2})+a^+(\vec{p},-\frac{1}{2})a^+(-\vec{p},-\frac{1}{2})\}\Psi_0\nonumber\\
&&-\frac{(\Lambda p)^0}{p^0}\sin\Omega_p\frac{1}{\sqrt{2}}\{
a^+(\vec{p},\frac{1}{2})a^+(-\vec{p},-\frac{1}{2})-a^+(\vec{p},-\frac{1}{2})a^+(-\vec{p},\frac{1}{2})\}\Psi_0\nonumber\\
&=&\frac{(\Lambda p)^0}{p^0}\{\cos\Omega_p \Psi'_{00} -
\sin\Omega_{p} \Psi'_{11}\},\label{ulpi}
\end{eqnarray}
where $\Psi'{ij}$ is the Bell states in the moving frame $S'$
whose momentums are transformed as $\vec{p}\to\vec{p}_{\Lambda},
-\vec{p}\to-\vec{p}_{\Lambda}$. Likewise, we have
\begin{eqnarray}
U(\Lambda)\Psi_{01}&=&\frac{(\Lambda p)^0}{p^0}
\Psi'_{01},\label{ulpii}\\
U(\Lambda)\Psi_{10}&=&\frac{(\Lambda p)^0}{p^0}
\Psi'_{10},\label{ulpiii},
\end{eqnarray}
and
\begin{equation}
U(\Lambda)\Psi_{11}=\frac{(\Lambda p)^0}{p^0}\{\sin\Omega_p
\Psi'_{00} +\cos\Omega_{p} \Psi'_{11}\}.\label{ulpiv}
\end{equation}
\end{mathletters}

If we regard $\Psi'_{ij}$ as Bell states
in the moving frame $S'$, then to an observer in $S'$, the effects
of the Lorentz transformation of the bipartite entangled Bell
states should appear as the superpositions of Bell states in the
frame $S'$.

The implications could be non trivial. One of the controversies in
modern physics is the violation of the local causality of
relativistic quantum field theory during the measurement
process\cite{tomm}. This is based on the EPR paradox\cite{eins}
and the Bell's theorem\cite{bell}, which suggest the existence of
nonlocal instantaneous action between distant measurements. In the
following, we investigate whether a supposed nonlocality is a real
physical property of the quantum theory, more specially, the
result of state collapse description by studying the case of an
entangled state shared by Alice and Bob in the relativistic
regime. For example, consider Alice in the frame $S$ and Bob in
the frame $S'$ (initially coincide with $S$) moving in the
$x$-direction share entangled pair of atoms whose electrons have
opposite momentum prepared at certain time $t=0$. At time $t=0$,
the entangled state shared by Alice and Bob is assumed to be
$\Psi^{AB}_{00}$,
\begin{eqnarray}
\Psi_{00}^{AB}&=&\Psi_{00}\nonumber\\
&=&\frac{1}{\sqrt{2}}\{ a^+_{A}(\vec{p},{{1}\over
{2}})a^+_{B}(-\vec{p},\frac{1}{2})+a^+_{A}(\vec{p},-\frac{1}{2})a^+_{B}(-\vec{p},-\frac{1}{2})\}\Psi_0.\label{bels}
\end{eqnarray}
Here A and B denote particles belong to Alice and Bob,
respectively. When the reference frame $S'$ where Bob is in, is
moving, the Lorentz boost $\Lambda$ will affect only the Alice's
state and as a result the global unitary transformation can be
written as
\begin{equation}
U_{AB}=U_{A}(\Lambda)\otimes I_B, \label{unita}
\end{equation}
where $U_{A}(\Lambda)$ is the unitary transformation representing
the Lorentz boost upon Alice. Then the quantum state from Bob's
point of view is given by
\begin{eqnarray}
U_{AB}(\Lambda)\Psi_{00}^{AB}&=&\frac{1}{\sqrt{2}}\sum_{\sigma}\sqrt{\frac{(\Lambda
p)^0}{p^0}}\{ {\cal
D}^{(\frac{1}{2})}_{\sigma\frac{1}{2}}(W(\Lambda,
p))a^+_{A}(\vec{p}_{\Lambda},\sigma)a^+_{B}(-\vec{p},\frac{1}{2})\nonumber\\
&&+{\cal D}^{(\frac{1}{2})}_{\sigma-\frac{1}{2}}(W(\Lambda,
p))a^+_{A}(\vec{p}_{\Lambda},\sigma)a^+_{B}(-\vec{p},-\frac{1}{2})\}\Psi_0\nonumber\\
&=&\sqrt{\frac{(\Lambda
p)^0}{p^0}}[\cos\frac{\Omega_p}{2}\frac{1}{\sqrt{2}}\{a^+_{A}(\vec{p}_{\Lambda},\frac{1}{2})
a^+_{B}(-\vec{p},\frac{1}{2})+a^+_{A}(\vec{p}_{\Lambda},-\frac{1}{2})
a^+_{B}(-\vec{p},-\frac{1}{2})\}\nonumber\\
&&-\sin\frac{\Omega_p}{2}\frac{1}{\sqrt{2}}\{a^+_{A}(\vec{p}_{\Lambda},\frac{1}{2})
a^+_{B}(-\vec{p},-\frac{1}{2})-a^+_{A}(\vec{p}_{\Lambda},-\frac{1}{2})
a^+_{B}(-\vec{p},\frac{1}{2})\}]\Psi_0\nonumber\\
&=&\sqrt{\frac{(\Lambda p)^0}{p^0}}[\cos\frac{\Omega_p}{2}
\Psi'^{AB}_{00} - \sin\frac{\Omega_{p}}{2}
\Psi'^{AB}_{11}],\label{ulpp}
\end{eqnarray}
Now Alice performs the measurement of the spin in the $+z$
direction at time $t=\tau$. Since the Bell state
$U_{AB}(\Lambda)\Psi^{AB}_{00}$ viewed from Bob in the frame $S'$
is a linear combination of $\Psi'^{AB}_{00}$ and
$\Psi'^{AB}_{11}$, when Alice measures her spin in the positive
$z$ direction, Bob's spin state is still a linear combination of
$|+\frac{1}{2}\rangle$ and $|-\frac{1}{2}\rangle$. This leaves
Bob's spin direction undetermined contradicting the EPR paradox.
On the other hand, in the non-relativistic quantum mechanics,
Bob's quantum state is determined instantaneously as a result of
collapse when Alice does her measurement which results in the
violation of the Bell inequality or the EPR paradox\cite{pres}.
In order to justify our argument, we have calculated the average of the Bell
observable
\begin{equation}
c(\vec{a}, \vec{a}', \vec{b}, \vec{b}')=<\hat{a} \otimes\hat{b}>
+<\hat{a} \otimes\hat{b}'>+<\hat{a}' \otimes\hat{b}>-<\hat{a}' \otimes\hat{b}'>
\label{bellin}
\end{equation}
where $\hat{a}, \hat{b}$ are the relativistic spin observables for Alice and Bob, respectively, related to
the Pauli-Lubanski pseudo vector which is known to be a relativistically invariant operator corresponding
to spin\cite{czac}\cite{ryde}--\cite{flemi}. Normalized relativistic spin
observables $\hat{a}, \hat{b}$ are given by\cite{czac},
\begin{equation}
\hat{a}=\frac{(\sqrt{1-\beta^2} \vec{a}_{\perp}+\vec{a}_{\parallel})\cdot\vec{\sigma}}{\sqrt{1+\beta^2[(\hat{n}\cdot\vec{a})-1]}}
\label{hata}
\end{equation}
and
\begin{equation}
\hat{b}=\vec{b}\cdot\vec{\sigma},
\end{equation}
where $\hat{n}$ is the direction of the Lorentz boost, $\beta=v/c$, $\vec{a}_{\perp}$ and $\vec{a}_{\parallel}$ are the components of
$\vec{a}$ which are perpendicular and parallel to the boost direction, respectively. Moreover, $|\vec{a}|=|\vec{b}|=1$.
Then, after some mathematical manipulations, we obtain
\begin{eqnarray}
<\hat{a}\otimes\hat{b}>=&&\frac{1}{\sqrt{1+\beta^2(a_x^2-1)}}[(b_x \cos\Omega_p+
b_z \sin\Omega_p)a_x-b_y a_y \sqrt{1-\beta^2}\nonumber\\
&&+\sqrt{1-\beta^2}(b_z \cos\Omega_p+ b_x \sin\Omega_p)a_z],
\label{expect}
\end{eqnarray}
for the state given by the equation (\ref{ulpp}).
We now consider the vector $\vec{a}=(\frac{1}{\sqrt{2}},-\frac{1}{\sqrt{2}},0),
\vec{a}'=(-\frac{1}{\sqrt{2}},-\frac{1}{\sqrt{2}},0), \vec{b}=(0, 1, 0), \vec{b}'=(1, 0, 0)$ which lead to the maximum
violation of the Bell's inequality in the non-relativistic domain, $\Omega_p=0$ and $\beta=0$. Then the Bell observable
for the $4$ relevant joint measurements becomes
\begin{eqnarray}
&&<\hat{a}\otimes\hat{b}>+<\hat{a}\otimes\hat{b}'>+<\hat{a}'\otimes\hat{b}>-<\hat{a}'\otimes\hat{b}'>\nonumber\\
&&=\frac{2}{\sqrt{2-\beta^2}}(\sqrt{1-\beta^2}+\cos\Omega_p)<2,
\label{bells}
\end{eqnarray}
in the ultra-relativistic limit where $\beta=1 $ and $\Omega_p\neq 0$.
So the relativistic entangled states (\ref{ulpp}), satisfies the Bell's inequality at least in the ultra-relativistic limit.
Moreover, in the limit $\beta\to 1$, the expectation value of the spin correlation $<\hat{a}\otimes\hat{b}>$ becomes
\begin{equation}
<\hat{a}\otimes\hat{b}>=\frac{a_x}{|a_x|}(b_x \cos\Omega_p+b_z \sin\Omega_p),
\label{epec}
\end{equation}
suggesting that Alice's and Bob's measurements are not correlated at all.
One might question the validity of our result since there are
experimental evidences\cite{clau}--\cite{mand} showing the
violation of Bell's inequality which suggests that reality is
nonlocal. It may be irrelevant whether the entanglement is maintained or not at ultra-relativistic limit
because Eq.(\ref{epec}) is the result of the transformation of both spin and Bell states.
So far the experiments that test Bell's inequality are
done with entangled photons which are massless, not with
spin-$\frac{1}{2}$ massive particles. It is interesting to note
that the representation of the Wigner's little group $W$ for the
massless particle is diagonal\cite{wein}, i.e.,
\begin{equation}
{\cal D}_{\sigma\sigma'}(W)=
\exp(i\theta\sigma)\delta_{\sigma'\sigma}, \label{ppunit}
\end{equation}
where $\theta$ is the angle related to the Lorentz boost
$\Lambda$. So the form of the entanglement is left invariant even
after the Lorentz boost and this will give the similar results as
in the case of non-relativistic quantum mechanics. However, for
massive particles, it doesn't look like that Nature favors
nonlocality suggested by the EPR paradox at least in the ultra-relativistic
limit if one wants to reconcile the principles of quantum
mechanics with those of special relativity.

In conclusion, we studied the Lorentz transformation of the
bipartite entangled quantum states explicitly and found that to an
observer in a moving frame, the Bell states appear as rotations
(or linear combination) of the Bell bases in that frame. It turns
out that the joint measurement of spin variables are not correlated
at the ultra-relativistic limit because the Lorentz transformation
of both spin and the Bell states.

\vspace{2.0cm}

\centerline{\bf Acknowledgements}

This work was supported by the Korean Ministry of Science and
Technology through the Creative Research Initiatives Program under
Contact No. M1-0116-00-0008. We are also indebted to prof. M. S. Kim for valuable discussions.

\end{document}